\documentclass[journal,11pt]{IEEEtran}

\usepackage{amsfonts}
\usepackage{stackengine}

\usepackage{amsfonts}
\usepackage{stackengine}

\usepackage{amsmath}
\usepackage{graphicx}
\usepackage{lipsum}
\usepackage{cite}
\usepackage{multirow}
\usepackage{amssymb}
\usepackage{latexsym}
\usepackage{textcomp}
\usepackage{colortbl}
\usepackage{array}
\usepackage{tabulary}
\usepackage{color}
\usepackage[labelsep=period]{caption}
\usepackage{adjustbox}	
\usepackage{booktabs}
\usepackage{siunitx}

\newtheorem{remark}{Remark}

\newtheorem{definition}{Definition}

\newcolumntype{C}[1]{>{\centering\arraybackslash}m{#1}}
\newcolumntype{R}[1]{>{\raggedleft\arraybackslash}m{#1}}
\newcolumntype{L}[1]{>{\raggedright\arraybackslash}m{#1}}
\newcolumntype{P}[1]{>{\centering\arraybackslash}p{#1}}
\newcolumntype{M}[1]{>{\centering\arraybackslash}m{#1}}

\begin{document}

\title{Self-Triggered  Coordination Control of Connected Automated Vehicles in Traffic Networks}


\author{Nader Meskin, Ehsan Sabouni,  Wei Xiao, and
        Christos G. Cassandras
       \thanks{This publication was made possible by the NPRP grant (No. 12S-0228-190177)	from the Qatar National Research Fund (a member of the Qatar Foundation).
	The statements made herein are solely the responsibility of the authors.}
\thanks{N. Meskin is with the Department of Electrical Engineering, Qatar University, Doha, Qatar,  \texttt{{\small nader.meskin@qu.edu.qa.}}}
\thanks{E. Sabouni and C. G Cassandras are with Boston University, Brookline, MA, \texttt{{\small \{esabouni, cgc\}@bu.edu}}}
\thanks{W. Xiao is with MIT, Cambridge, MA, \texttt{{\small weixy@mit.edu}}}
}



\maketitle

\begin{abstract}
In this paper, a self-triggered scheme is proposed to optimally control the traffic flow of Connected and Automated Vehicles (CAVs)  at conflict areas of a  traffic network with the main aim of reducing the data exchange among CAVs in the control zone and at the same to  minimize the travel time and energy consumption. The safety constraints and the vehicle limitations are considered using the Control Barrier Function (CBF) framework and a self-triggered scheme is proposed using the CBF constraints. Moreover, modified CBF constraints are developed to ensure a minimum inter-event interval for the proposed self-triggered schemes. Finally, it is shown through  a simulation study that the number of data exchanges among CAVs is significantly reduced using the proposed self-triggered schemes in comparison with the standard time-triggered framework.
%
%
\end{abstract}
%
%
%
%

\section{Introduction}

The emergence of Connected Autonomous Vehicles (CAVs) and recent advances in intelligent transportation system technologies \cite{6587095} can potentially enhance a traffic network's performance with the ultimate aim of reducing energy consumption, air pollution, congestion, and at the same time to enhance safety. 
Traffic management at conflict areas, such as merging points in highway on-ramps, is one of the major challenges for future intelligent transportation systems in which safety, congestion, comfort, and energy consumption should be considered \cite{Waard2009}. 

There have been centralized and decentralized control algorithms developed for motion planning and coordination control of CAVs in a merging area \cite{6587095}. An approach is called ``centralized'' if  at least one task in the system is globally decided for all vehicles by a single central controller \cite{rajamani2000demonstration,shladover1991automated,xu2019grouping}. Such approaches work well when the safety constraint is independent of speed, but tend to be conservative and lack robustness to disturbances \cite{xiao2021decentralized}. In decentralized approaches, each CAV is controlled as an autonomous agent with the main objective of maximizing its own efficiency in the presence of safety constraints. Several decentralized merging control mechanisms have been developed in the literature \cite{malikopoulos2018decentralized,milanes2010automated,raravi2007merge,rios2016automated,xiao2021decentralized}. Specifically, in a decentralized optimal control framework, several optimization objectives such as the minimization of acceleration  \cite{rios2016automated}, the maximization of passenger comfort \cite{ntousakis2016optimal,rathgeber2015optimal}, or travel time of each CAV along with energy consumption \cite{xiao2021decentralized} are considered.  Model Predictive Control (MPC) techniques are  also developed to account for additional constraints \cite{cao2015cooperative,mukai2017model,nor2018merging,ntousakis2016optimal}. As an alternative to MPC, methods based on Control Barrier Functions (CBFs) are recently proposed in \cite{xiao2021bridging} where a joint optimal control and CBF function  controller (termed OCBF) is designed to account for optimality, safety, and computational complexity.
Unlike MPC, which is effective for problems with simple dynamics, objectives and constraints, the CBF-based method maps any continuously differentiable state constraint onto a new constraint on the control input and can guarantee forward invariance of the associated set by solving a sequence of quadratic programming (QP) problems. This allows the CBF method to be more effective for complex objectives, nonlinear dynamics, and constraints \cite{xu2020general}. 

It should be noted that in all previous work to date \cite{malikopoulos2018decentralized,rios2016automated,xiao2021decentralized,cao2015cooperative,mukai2017model,nor2018merging,ntousakis2016optimal,xiao2021bridging}, time-triggered communication between CAVs and the coordinator is assumed, i.e., all vehicles send their information to the coordinator at each time instant. It is clear that imposing such simultaneous time-triggered communication of CAVs is indeed very restrictive and practically not feasible. 
In this paper, a self-triggered coordination scheme is proposed where CAVs are communicating with the coordinator asynchronously such that at each self-triggered time instant CAV information is updated at the coordinator and a CAV also downloads any required state information from other CAVs.  The key advantage is to reduce the communication rate and also eliminate the need to synchronize communication between all CAVs and the coordinator. Toward this goal, first a set of \emph{modified} CBF constraints is obtained to ensure a minimum inter-event time interval. Then, all update time instants are obtained by calculating the first time instant that any of the safety CBF constraints is violated. Finally, it is shown that the communication rate between the CAVs and the coordinator is significantly reduced in comparison with the time-triggered scheme.

This paper is organized as follows. Section \ref{Sec_Pre} provides the preliminary definitions. In Section \ref{Sec_Problem}, the problem formulation is presented and in Section \ref{Sec_Solution}, the proposed self-triggered scheme is introduced. Simulation results are given in Section \ref{Sec_sim} and finally Section \ref{Sec_Con} concludes the paper.
\section{Preliminaries} \label{Sec_Pre}

Consider a control affine system
\begin{align}\label{eq_nonlinear}
    \dot{x}=f(x)+g(x)u,
\end{align}
where $f:\mathbb{R}^n \rightarrow \mathbb{R}^n$ and $g:\mathbb{R}^n \rightarrow \mathbb{R}^{n\times q}$ are locally Lipschitz, $x\in \mathbb{R}^n$ denotes the state vector and $u\in U\subset \mathbb{R}^q$ with $U$ as the control input constraint set. It is assumed that the solution of \eqref{eq_nonlinear} is forward complete.

\begin{definition}
A set $\mathcal{C}$ is forward invariant for system \eqref{eq_nonlinear} if for every $x(0) \in \mathcal{C}$, we have $x(t) \in \mathcal{C}$, for all $t>0$.
\end{definition}
\begin{definition}\cite{7782377}\label{Def_3}
Let $\mathcal{C}:=\{x\in \mathbb{R}^n:h(x)\geq 0\}$ with a continuously differentiable function $h:\mathbb{R}^n\rightarrow \mathbb{R}$. The function $h$ is a called Control Barrier Function (CBF) defined on set $\mathcal{D} \subset \mathcal{C} \subset \mathbb{R}^n$, if there exists a extended class $\mathcal{K}$ function $\alpha$ such that $\sup_{u \in U}[L_fh(x)+L_gh(x)u+\alpha(h(x))]\geq 0,~~\forall x\in \mathcal{D}$, where $L_f$, $L_g$ denote the Lie derivatives along $f$ and $g$, respectively.
\end{definition}

\begin{definition}\cite{6426229}
A continuously differentiable function $V:\mathbb{R}^n \rightarrow \mathbb{R}$ is a globally and exponentially stabilizing Control Lyapunov function (CLF) for \eqref{eq_nonlinear} if there exist constant $c_i>0$, $i=1,2,3$ such that $c_1 ||x||^2 \leq V(x) \leq c_2 ||x||^2$, and $\inf_{u\in U} [L_fV(x)+L_gV(x)u+c_3V(x)]\leq 0$.
\end{definition}

\section{Problem Formulation}\label{Sec_Problem}
\begin{figure}
	\centering
	\includegraphics[scale= 0.93]{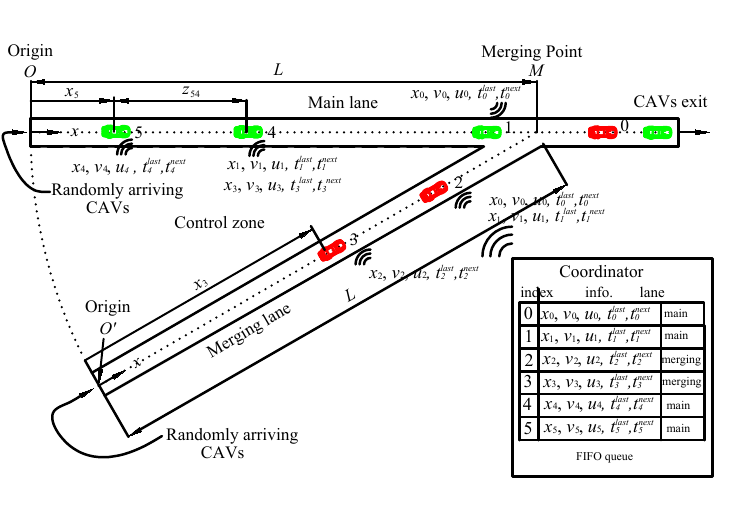}
	\caption{The merging problem.}
	\label{Fig:Merge_Problem}
\end{figure}

In this section, the cooperative motion control of CAVs  is reviewed at conflict areas of a traffic
network such as merging roads, signal-free intersections,
roundabouts, and highway segments where lane change
maneuvers take place.  A Control Zone (CZ) is defined as
an area within which CAVs can communicate with each other
or with a coordinator (e.g., a Road-Side Unit (RSU)) which
is responsible for facilitating the exchange of information
(but not control individual vehicles) within this CZ. As an
example, Fig. \ref{Fig:Merge_Problem} shows a conflict area due to vehicles merging
from two single-lane roads and there is a single Merging
Point (MP) which vehicles must cross from either road \cite{xiao2021decentralized}.
More generally, the CZ may include a set of MPs that each
CAV has to cross; for instance in a 4-way intersection with
two lanes per direction there are 32 total MPs \cite{xu2020general}.


Let $S(t)$ be the set of CAV  indices located in the CZ at time $t$ including the CAV with index 0 that just left the CZ with the cardinality of $N(t)$. Hence, the index of the next arriving CAV will have assigned as $N(t)$. Once a vehicle leaves the CZ, its index will be dropped from $S(t)$ and all the remaining indices are decreased by one. 

The vehicle dynamics of the $i$-th CAV, $i\in S(t)$ along  its dedicated lane is given as 
\begin{align}\label{CAV_Dynamics}
\dot{x}_i(t)&=v_i(t),~~~\dot{v}_i(t)=u_i(t),    
\end{align}
where $x_i(t)$ is the distance to the origin of the lane at which the $i$-th CAV arrives, and $v_i(t)$ and $u_i(t)$ are the velocity and the acceleration control input of the $i$-th CAV, respectively.

\begin{remark}
It should be noted that it is assumed that the local controller implemented on each CAV is taking care of the nonlinear longitudinal model and hence from the coordination perspective, one can model the CAVs as in \eqref{CAV_Dynamics}. 
\end{remark}

The following objectives and constraints are considered for the safe and optimal coordination of CAVs at a given conflict zone.

\noindent \textbf{Objective 1} (Minimizing travel time): Let $t_i^0$ and $t_i^f$ denote the time that the $i$-th CAV arrives at the origin and leaves the CZ, respectively. The first objective is to minimize the travel time $t_i^f-t_i^0$ for all $i\in S(t)$.\\
\textbf{Objective 2} (Minimizing energy consumption): We also aim to minimize the energy consumption for each CAV $i \in S(t)$ expressed as
\begin{align*}
J_i(u_i(t))=\int_{t_i^0}^{t_i^f} C(u_i(\tau)) \textmd{d}\tau,
\end{align*}
where $C(\cdot)$ is a strictly increasing function of its argument. In order to minimize the energy consumption, one can select the cost function $C(u_i(t))=\frac{1}{2}u_i^2(t)$. \\

\noindent \textbf{Constraint 1} (Safety):
Let $i_p$ denote the index of the CAV that immediately precedes the $i$-th CAV in the same lane in the CZ (if one is present). It is required that the distance between the center of $i$-th CAV and the center of $i_p$-th CAV, denoted as $z_{i,i_p}(t)=x_{i_p}(t)-x_i(t)$, be constrained by the velocity $v_i(t)$ of the $i$-th CAV such that
\begin{align} \label{h1}
z_{i,i_p}(t) \geq \psi v_i(t)+l,~\forall t\in [t_i^0,t_i^f],
\end{align}
where $\psi$ denotes the reaction time generally selected as $\psi=1.8$ \cite{VOGEL2003427} and $l$ is a constant which depends on the length of these two CAVs.\\
\noindent \textbf{Constraint 2} (Safe Merging): In order to avoid collision at MPs, for each CAV $i\in S(t)$, the following safe margin distance should be imposed at $t_i^{M_k}$,
\begin{align}\label{safe_merging}
z_{i,j}(t_i^{M_k}) \geq \psi v_i(t_i^{M_k}  )+l,
\end{align}
where the index $j$ refers to the CAV that may collide with the $i$-th CAV at the merging point $k$, $k\in\{1,\dots,n_i\}$, with  $n_i$ as the
total number of MPs that the $i$-th CAV passes in the CZ and $t_i^{M_k}$ as the corresponding time instant of passing the $k$-th MP. Based on the policy adopted for
sequencing CAVs through the CZ, CAV $j$ is determined. Different approaches  such as First-In-First-Out
(FIFO) based on the arrival times of CAVs, the Dynamic
Resequencing (DR) policy  or any other
desired policy can be adopted for CAV sequencing.

It should be noted  that this constraint is only
applied at a certain time instant $t^{M_k}_i$ which obviously depends on how
the CAVs are controlled. As an example, in Fig. 1 under
FIFO, we have $j = i-1$ and $t^{M_k}_i=t_i^f$ since the MP defines
the exit from the CZ. As explained in \cite{10.1145/3302509.3311054}, in order to apply the CBF approach, it is required to have a continuously differentiable version of the above constraint and one feasible example is given as
\begin{align}\label{h2}
z_{i,j}(t) \geq\Phi(x_i(t))v_i(t)+l,~\forall t\in [t_i^0,t_i^{M_k}],
\end{align}
where $\Phi:\mathbb{R}\rightarrow \mathbb{R}$ can be any continuously differentiable function as long as it is strictly increasing with the following boundary conditions: $\Phi(x_i(t_0))=0$ and $\Phi(x_i(t_i^{M_k}))=\psi$. One possible choice is a linear function $\Phi(x_i(t))=\psi \frac{x_i(t)}{L}$ where $L$ is the length of the road from the origin $O$ or $O'$ to the MP.\\
\noindent \textbf{Constraint 3} (Vehicle limitation): There exist constraints on the speed and acceleration of each CAV in the CZ as follows:
\begin{align}\label{h3}
v_{\textmd{min}}\leq v_i(t) \leq v_{\textmd{max}},~\forall t\in [t_i^0,t_i^f],\\ \label{h4}
u_{\textmd{min}}\leq u_i(t)\leq  u_{\textmd{max}} ,~\forall t\in [t_i^0,t_i^f],
\end{align}
where $v_{\max}>0$ and $v_{\min}\geq0$ denote the maximum and minimum allowable velocity in the road, $u_{\textmd{min}}=-c_d  g$, $u_{\textmd{max}}=c_ag$,  $c_d>0$ and $c_a>0$ are deceleration and acceleration coefficients, respectively, and $g$ is the gravity constant.

Considering both of the above objectives, the cost function for the $i$-th CAV can be written as:
\begin{align}
    J_i(u_i(t))=\int_{t_i^0}^{t_i^f} \big (\alpha +\frac{(1-\alpha)\frac{1}{2}u_i^2(\tau)}{\frac{1}{2} \max\{u_{\max}^2,u_{\min}^2\}}\big)\textmd{d} \tau,
\end{align}
where $\alpha \in [0,1]$. For $\alpha<1$, we can define $\beta=\frac{\alpha \max\{u_{\max}^2,u_{\min}^2\}}{2(1-\alpha)}$ and consequently, the cost function is given as 
\begin{align}\label{cost}
    J_i(u_i(t))=\beta (t_i^f-t_i^0)+ \int_{t_i^0}^{t_i^f} \frac{1}{2}u_i^2(\tau)\textmd{d} \tau,
\end{align}
where $\beta$ is a weight factor for combining the travel time and energy consumption minimization objectives.

\textbf{Problem 1:} The main goal of this paper  is to determine the control laws to achieve Objectives 1, 2 subject to Constraints 1, 2, 3 for each CAV in the control zone with the dynamics \eqref{CAV_Dynamics} using a self-triggered communication framework for data exchange among the vehicles.   

As already mentioned, in all previous works \cite{10.1145/3302509.3311054,xiao2021decentralized,xiao2021bridging}, time-triggered communication between CAVs and the coordinator is assumed, i.e., all vehicles send their state information to the coordinator at each time instant. It is clear that imposing such synchronous time-triggered communication of CAVs in the CZ is indeed very restrictive and practically not feasible. To remedy this issue, either a self-triggered or an event-triggered scheme can be adopted. In
\cite{event_triggered}, an event-triggered scheme is proposed such that  CAV $i$ updates its control input whenever its states or states of CAV $j$ or CAV $i_p$ reach a given adjustable bound and it is shown that the safety constraints can be guaranteed, while the number of times that communication is required is significantly reduced.
 In this paper, a self-triggered asynchronous communication scheme is considered where each CAV $i$ communicates with the coordinator at specified time instants $\{t_i^k\}_{k\in \mathbb{Z}^+}$. At each such instant $t_i^k$, CAV $i$ uploads its own state information as $x_i(t_i^k)$, $v_i(t_i^k)$, the calculated control input $u_i(t_i^k)$ that is going to be applied over the time interval $[t_i^k,t_i^{k+1})$, and the next time CAV $i$ will communicate with the coordinator, denoted as $t_i^{\textmd{next}}$. The data stored at the coordinator for all vehicles are shown in Table \ref{tab:1}. We denote the latest stored information of the $i$-th CAV at the coordinator as $\mathcal{I}_i=[t_i^{\textmd{last}},t_i^{\textmd{next}},x_i(t_i^{\textmd{last}}),v_i(t_i^{\textmd{last}}),u_i(t_i^{\textmd{last}})]
$. 
\begin{table}
\begin{center}
\begin{tabular}{|ll|}
\hline
$t_i^{\textmd{last}}$ & Last time CAV $i$ communicated.\\
$t_i^{\textmd{next}}$ & Next time CAV $i$ will communicate.\\
$x_i(t_i^{\textmd{last}})$ & Last updated position of CAV $i$.\\
$v_i(t_i^{\textmd{last}})$ & Last updated velocity of CAV $i$.\\
 $u_i(t_i^{\textmd{last}})$ & Last control input of CAV $i$.\\
 \hline
\end{tabular}
\caption{Data stored on the coordinator for all vehicles}\label{tab:1}
\end{center}
\end{table}

The main aim is to develop a self-triggered asynchronous algorithm to determine the sequence of time instants  $\{t_i^k\}_{k\in \mathbb{Z}^+}$ and the control input $u_i(t), t\in [t_i^k,t_i^{k+1}), k\in \mathbb{Z}^+$ for each CAV to solve Problem 1. One important feature in a self-triggered scheme is to guarantee a lower bound for the inter-event time interval. In other words, for the generated time instants  $\{t_i^k\}_{k\in \mathbb{Z}^+}$, there should exist some $T_d>0$  such that  $|t_i^{k+1}-t_i^k|\geq T_d$,  $\forall k\in \mathbb{Z}^+$. This is a design parameter which depends on the sensor sampling rate, as well as the clock of the on-board embedded system on each CAV. For the same reason, the time-instants  $\{t_i^k\}_{k\in \mathbb{Z}^+}$ are calculated such that $(t_i^k~\textrm{mod}~T_d)=0$ where $\textrm{mod}$ denotes the modulo operator.


\section{Proposed Solution}\label{Sec_Solution}

Based on the CBF framework, in order to satisfy the safety constraints \eqref{h1} and \eqref{h2}, as well as the vehicle state and control limitations \eqref{h3} and \eqref{h4}, as per Definition \ref{Def_3} the following CBFs are defined for the $i$-th CAV:
\begin{align}\label{CBF1}
h_{i,1}(t)=&v_{\textmd{max}}-v_i(t),\\ \label{CBF2}
h_{i,2}(t)=&v_i(t)-v_{\textmd{min}},\\ \label{CBF3}
h_{i,3}(t)=&x_{i_p}(t)-x_i(t)-\psi v_i(t)-l,\\ 
h_{i,4}(t)=&x_{j}(t)-x_i(t)- \frac{\psi x_i(t)v_i(t)}{L}-l.\label{CBF4}
\end{align}
Moreover, as shown in \cite{xiao2021bridging}, the unconstrained optimal solution for minimizing the cost function \eqref{cost} the $i$-th CAV is given as $ u_i^*(t)=a_i t+b_i$, $v_i^*(t)=\frac{1}{2}a_i t^2+b_i t+ c_i$
where the coefficient $a_i$, $b_i$, and $c_i$ can be obtained by solving a set of nonlinear algebraic equations as detailed in \cite{xiao2021bridging}.  Hence, the solution of Problem 1 determines a controller for each CAV such that it can track the above optimal solutions as close as possible and at the same to ensure Constraints 1, 2, and 3. Toward this, the following optional control Lyapunov function is selected as:
\begin{align}\label{CLF}
V(v_i(t),v_i^*(t))=\frac{1}{2}(v_i(t)-v_i^*(t))^2,
\end{align}
to further enforce the optimal solution obtained from the unconstrained optimal solution. In previous works \cite{10.1145/3302509.3311054,xiao2021decentralized,xiao2021bridging},  Problem  1 is solved in a time-triggered fashion  with a fixed sampling time $T_s$ for all CAVs, 
 where $[t_i^0,t_i^f]$ is divided into intervals $[t_i^0,t_i^0+T_s],...,[t_i^0+kT_s,t_i^0+(k+1)T_s],...$ and then, based on the defined CBFs  in \eqref{CBF1}-\eqref{CBF4}, and the Lyapunov function \eqref{CLF}, the following sequence of quadratic programming (QP) problems can be formalized at each time instant $t=t_i^0+kT_s$, $k=0,1,\dots$, to solve Problem 1, namely:
\begin{align}  \label{QP}
\min_{u_i(t),\delta_i(t)}~~\frac{1}{2}(u_i(t)-u_i^*(t))^2+\rho \delta_i^2(t)
\end{align}
subject to
\begin{align}\label{eq_11}
(v_i(t)-v_i^*&(t))u_i(t)+c_3(v_i(t)-v_i^*(t))^2\leq \delta_i(t),\\  \label{eq_12} 
\mathcal{C}_{i,1}(t,u_i(t))&=-u_i(t)+\alpha_1(h_{i,1}(t))\geq 0,\\\label{eq_13}
\mathcal{C}_{i,2}(t,u_i(t))&=u_i(t)+\alpha_2(h_{i,2}(t)) \geq 0,\\ \label{eq_14}
\mathcal{C}_{i,3}(t,u_i(t))&=v_{i_p}(t)-v_i(t)-{\psi} u_i(t)+\alpha_3(h_{i,3}(t)) \geq 0,\\
\nonumber
\mathcal{C}_{i,4}(t,u_i(t))&=v_{j}(t)-v_i(t) -\frac{\psi}{L}v_i^2(t)-\frac{\psi}{L}x_i(t)u_i(t)\\ \label{eq_15} &+\alpha_4(h_{i,4}(t))>0,\\
u_{\textmd{min}}&\leq u_i(t)\leq  u_{\textmd{max}} ,
\end{align}
where $\delta_i(t)$ is a relaxation variable that makes the control Lyapunov function constraint \eqref{CLF} a soft constraint. It should be noted that the decision variables $u_i(t)$ and $\delta_i(t)$ are assumed to be constant in each time interval.


In this paper, in contrast to the time-triggered scheme with a fixed sampling time $T_s$, each CAV $i \in S(t)$ calculates the time instant $t_i^k$ in which the  QP problem should be solved in a self-triggered fashion. This is mainly achieved  such that  at each time instant $t_i^k$,  CAV $i$ solves its QP problem to obtain $u_i(t_i^k)$ and also calculates the next time instant $t_i^{k+1}$ in which it should solve the QP problem  and as in the time-triggered scheme, the calculated fixed control input $u_i(t_i^k)$ (constant acceleration) is applied over the time interval  $[t_i^k,t_i^{k+1})$. 

In view of the constraints \eqref{eq_14} and \eqref{eq_15}, CAV $i$ requires knowledge of $v_{i_p}(t_i^k)$, $x_{i_p}(t_i^k)$, $v_{j}(t_i^k)$, and $x_{j}(t_i^k)$ at time instant $t_i^k$. Hence, at each time instant that it accesses the coordinator, it needs to download the recorded data of CAV $i_p$ and $j$, namely, $\mathcal{I}_{i_p}$ and $\mathcal{I}_{j}$. Then, the required updated information at  $t_i^k$ for CAV $i_p$ can be calculated as
$v_{i_p}(t_i^k)= v_{i_p}(t_{i_p}^\textmd{last})+(t_i^k-t_{i_p}^\textmd{last})a_{i_p}(t_{i_p}^\textmd{last})$,
$x_{i_p}(t_i^k)= x_{i_p}(t_{i_p}^\textmd{last})+(t_i^k-t_{i_p}^\textmd{last})v_{i_p}(t_{i_p}^\textmd{last})+\frac{1}{2}(t_i^k-t_{i_p}^\textmd{last})^2u_{i_p}(t_{i_p}^\textmd{last})$ with similar information calculated for CAV $j$. The information for CAV $i_p$ may also be obtained from the on-board sensors in CAV $i$.



Therefore, the remaining problem to be solved is how each CAV $i \in S(t)$ should specify the time instants $t_i^k$, $\forall k\in \mathbb{Z}^+$. First, it will be shown how a lower bound $T_d$ on the inter-event time interval can be ensured.

\subsection{Minimum Inter-event Time}
In this subsection, it is shown how the CBF constraints \eqref{eq_12} to \eqref{eq_15} should be modified to ensure a minimum inter-event time $T_d$. This is achieved by adding extra positive terms to the right hand side of these constraints.  For simplicity, the functions $\alpha_i(r)$ are selected as $\alpha_i(r)=r,i=1,\dots,4$. First, consider the maximum speed CBF constraint \eqref{eq_12} and solving the QP problem at $t_i^k$ with feasible solution $u_i(t_i^k)$. It follows that:
\begin{align} \label{cbf_i1}
\mathcal{C}_{i,1}(t_i^k,u_i(t_i^k))&=-u_i(t_i^k)+v_{\textmd{max}}-v_i(t_i^k)\geq 0.
\end{align}
However, the CBF constraint should be satisfied in the time interval $[t_i^k,t_i^k+T_d]$, i.e.
\begin{align} 
\mathcal{C}_{i,1}(t,u_i(t_i^k))=-u_i(t_i^k)+&v_{\textmd{max}}-v_i(t)\geq 0,\label{eq_c1}
\end{align}
$\forall t\in [t_i^k,t_i^k+T_d]$, and it follows from $v_i(t)=v_i(t_i^k)+u_i(t_i^k)\tau$  that
\begin{align} \nonumber
\mathcal{C}_{i,1}(t,u_i(t_i^k))&=\mathcal{C}_{i,1}(t_i^k,u_i(t_i^k))-u_i(t_i^k)\tau, \tau \in [0,T_d],
\end{align}
where $\tau=t-t_i^k$. Hence, the difference between the CBF constraints \eqref{cbf_i1} and \eqref{eq_c1} is given as
\begin{align*}
\mathcal{M}_{i,1}&(t,t_i^k,u_i(t_i^k)):=  \mathcal{C}_{i,1}(t_i^k,u_i(t_i^k))-\mathcal{C}_{i,1}(t,u_i(t_i^k))\\ \leq &u_{M}\tau:=\nu_{i,1}(\tau),
\end{align*}
where $u_M=\max(u_{\min},u_{\max})$. Therefore, in order to enforce \eqref{eq_c1}, the CBF constraint should be modified as 
$\mathcal{C}_{i,1}(t_i^k,u_i(t_i^k))\geq \nu_{i,1}(T_d)$. Then, it follows that 
\begin{align*}
 \mathcal{C}_{i,1}(t_i^k,u_i(t_i^k))&- \mathcal{C}_{i,1}(t,u_i(t_i^k))\\&+  \mathcal{C}_{i,1}(t,u_i(t_i^k))\geq \nu_{i,1}(T_d),
\end{align*}
which leads to $    \mathcal{C}_{i,1}(t,u_i(t_i^k))\geq \nu_{i,1}(T_d) -\mathcal{M}_{i,1}(t,t_i^k,u_i(t_i^k))\geq 0,$ for $t \in [t_i^k,t_i^k+T_d]$. Hence, in order to ensure the minimum inter-event interval $T_d$, the CBF constraint \eqref{eq_12} should be modified to 
\begin{align}\label{cbf_modified_1}
    -u_i(t)+v_{\textmd{max}}-v_i(t)\geq \nu_{i,1}(T_d).
\end{align}
Following a similar derivation, the minimum speed CBF constraint \eqref{eq_13} should be modified to  
\begin{align}\label{cbf_modified_2}
  u_i(t)+v_i(t)-v_{\textmd{min}} \geq \nu_{i,2}(T_d).
\end{align}
where $\nu_{i,2}(T_d)=u_{M}T_d$.

Next, let us consider the safety CBF constraint \eqref{eq_14} and solving the QP problem at $t_i^k$ with a feasible solution $u_i(t_i^k)$. It follows that 
\begin{align} \nonumber
  \mathcal{C}_{i,3}(t_i^k,u_i(t_i^k))= & v_{i_p}(t_i^k)-v_i(t_i^k)-{\psi} u_i(t_i^k)\\ &+h_{i,3}(t_i^k)\geq 0. \label{eq_c2}
\end{align}
However, we need to ensure at least that 
\begin{align} \nonumber
     \mathcal{C}_{i,3}(t,u_i(t_i^k))= &v_{i_p}(t)-v_i(t)-{\psi} u_i(t_i^k)\\&+h_{i,3}(t) \geq 0, t \in [t_i^k,t_i^k+T_d], \label{eq_30}
\end{align}
and it follows from  
$ v_{i_p}(t)-v_i(t)
= \Delta v_{i,i_p}(t_i^k)+\Delta u_{i,i_p}(t_i^k)\tau $
 that
\begin{align}\nonumber
 \mathcal{C}_{i,3}(t,u_i(t_i^k))= & \mathcal{C}_{i,3}(t_i^k,u_i(t_i^k))+\Delta u_{i,i_p}(t_i^k)\tau \\ \nonumber &+ \Delta v_{i,i_p}(t_i^k)\tau+0.5 \tau^2 \Delta u_{i,i_p}(t_i^k)\\&-\psi u_i(t_i^k)\tau \geq 0, \tau \in [0,T_d] \label{Ci3}.
\end{align}
where $\tau=t-t_i^k$, $\Delta v_{i,i_p}(t_i^k)=v_{i_p}(t_{i}^k)-v_i(t_i^k)$ and $\Delta u_{i,i_p}(t_i^k)=u_{i_p}(t_{i}^k)-u_i(t_i^k)$. Hence, the difference between the CBF constraints \eqref{eq_c2} and \eqref{eq_30} is given as
\begin{align*}
\mathcal{M}_{i,3}&(t,t_i^k,u_i(t_i^k)):=  \mathcal{C}_{i,3}(t_i^k,u_i(t_i^k))-\mathcal{C}_{i,3}(t,u_i(t_i^k))\\ \leq &(|u_{i_p}(t_i^k)|+(1+\psi)u_M+|v_{i,i_p}(t_i^k)|)\tau\\&+0.5 \tau^2 (|u_{i_p}(t_i^k)|+u_M):=\nu_{i,3}(\tau,t_i^k).
\end{align*}
Therefore, by modifying the CBF constraint \eqref{eq_14} to 
\begin{align} \label{cbf_modified_3}
\mathcal{C}_{i,3}(t,u_i(t))\geq  \nu_{i,3}(T_d,t),
\end{align}
one can enforce \eqref{eq_30}. Indeed, if the above modified CBF constraint is satisfied at time-instant $t_i^k$, i.e.
 $\mathcal{C}_{i,3}(t_i^k,u_i(t_i^k))\geq \nu_{i,3}(T_d,t_i^k)$
then, it follows that:
\begin{align*}
    \mathcal{C}_{i,3}(t,u_i(t_i^k))\geq &\nu_{i,3}(T_d,t_i^k)-\mathcal{M}_{i,3}(t,t_i^k,u_i(t_i^k))\geq 0
\end{align*}
for $t \in [t_i^k,t_i^k+T_d]$. 

Following a similar approach, it can be shown that the CBF constraint \eqref{eq_15} should be modified to 
\begin{align} \label{cbf_modified_4}
  \mathcal{C}_{i,4}(t,u_i(t))\geq \nu_{i,4}(T_d,t),
\end{align}
where  $ \nu_{i,4}(T_d,t)=\Big(|u_{j}(t)|+ ( \frac{3\psi}{L}|v_i(t)|+\frac{\psi}{L}|x_i(t)|+1)u_M+|v_{j}(t)|+|v_i(t)|+\frac{\psi}{L}v_i^2(t) \Big) T_d +\Big( \frac{3\psi}{2L}u_M^2+0.5 |u_{j}(t)|+0.5u_M+\frac{3\psi}{2L}  |v_i(t)| u_M \Big)T^2_d+0.5*\frac{\psi}{L}u^2_M T_d^3.$

Finally, as the CLF constraint \eqref{eq_11} is mainly added optionally for following the optimal trajectory and it can be relaxed in the presence of safety constraint, there is generally no need to assure that it is satisfied during the whole time-interval $t \in [t_i^k,t_i^k+T_d]$ with the same relaxing variable value $\delta_i(t_i^k)$ and hence it is not needed to be modified as for CBF constraints. Therefore,  in order to ensure the minimum inter-event time $T_d$, at each time instant $t_i^k$, CAV $i$ needs to solve the following QP
\begin{align}  \label{QP2}
 \min_{u_i,\delta_i}~~\frac{1}{2}(u_i-u_i^*(t_i^k))^2+\rho \delta_i^2
\end{align}
subject to the modified CBF constraints \eqref{cbf_modified_1}, \eqref{cbf_modified_2}, \eqref{cbf_modified_3}, and \eqref{cbf_modified_4}, the control input bounds \eqref{h4} and the CLF constraint \eqref{eq_11}. In the next subsection, it will be shown how the time-instant $t_i^k$ should be obtained for each CAV.

\subsection{Self-Triggered Time Instant Calculation}

The key idea in this self-triggered framework is to predict the first time instant that any of the CBF constraints \eqref{eq_12}, \eqref{eq_13}, \eqref{eq_14}, \eqref{eq_15} is violated and select that as the next time instant $t_i^{k+1}$ to communicate with the coordinator and to update the control input action. We point out that it is not required to consider the modified CBF constraints \eqref{cbf_modified_1}, \eqref{cbf_modified_2}, \eqref{cbf_modified_3}, and \eqref{cbf_modified_4} here, as they are obtained for ensuring the minimum inter-event time $T_d$, while the original CBF constraints are sufficient for satisfying Constraints 1, 2, and 3 in Problem 1.

For the constraint \eqref{eq_12},  it is clear that if $u_i(t_i^k)\leq0$ (decelerating), then this constraint is satisfied, hence there is no need to check it. However, for $u_i(t_i^k) >0$, the constraint \eqref{eq_12} is violated if
$-u_i(t_i^k)+v_{\textmd{max}}-v_i(t_i^k)-u_i(t_i^k) (t-t_i^k)= 0$
which leads to the time instant 
\begin{align*}
t_{i,1}^k=t_i^k+\frac{-u_i(t_i^k)+v_{\textmd{max}}-v_i(t_i^k)}{u_i(t_i^k)}.
\end{align*}


{Observe that as at the time instant $t_i^k$, the QP \eqref{QP2} is solved and it follows that the constraint \eqref{cbf_modified_1} is satisfied at $t=t_i^k$ and we have $-u_i(t_i^k)+v_{\textmd{max}}-v_i(t_i^k)  \geq \nu_{i,1}(T_d) > 0$; therefore, $t^k_{i,1}\geq t_i^k+T_d$. }

For the constraint \eqref{eq_13},  it is clear that if $u_i(t_i^k)\geq0$ (accelerating), then this constraint is satisfied and hence no need to check this one. However, for $u_i(t_i^k) <0$, the constraint \eqref{eq_13} is violated if $u_i(t_i^k)+v_i(t_i^k)+u_i(t_i^k) (t-t_i^k)-v_{\textmd{min}} = 0$ which leads to the time instant 
\begin{align*}
t^k_{i,2}=t_i^k+\frac{-u_i(t_i^k)+v_{\textmd{min}}-v_i(t_i^k)}{u_i(t_i^k)},
\end{align*}
and it can be shown similar to the previous case that $t^k_{i,2}\geq t_i^k+T_d$.

For the constraint \eqref{eq_14},  we need to find the first time instant $t>t_i^k$ such that $\mathcal{C}_{i,3}(t,u_i(t_i^k))=0$. Based on 
 \eqref{Ci3}, this leads to the following quadratic equation
\begin{align*}
&\big (0.5  \Delta u_{i,i_p}(t_i^k)\big)\tau^2+\big ( \Delta u_{i,i_p}(t_i^k)+(\Delta v_{i,i_p}(t_i^k)\\&-\psi u_i(t_i^k) )\big )\tau+ \mathcal{C}_{i,3}(t_i^k,u_i(t_i^k))=0.
\end{align*}
The least positive root of the above equation is denoted  as $\tau_{i,3}$ and we define $t^k_{i,3}=t_i^k+\tau_{i,3}$. { The case of having both roots negative corresponds to the scenario that the constraint \eqref{eq_14} will not be violated, hence $t^k_{i,3}=\infty$. } Moreover, due to the added term in \eqref{cbf_modified_3}, it follows that $\tau_{i,3}\geq T_d$.

Similarly for the constraint \eqref{eq_15},  the first time instant  $t>t_i^k$ such that $\mathcal{C}_{i,4}(t,u_i(t_i^k))=0$ can be obtained by solving the following cubic equation 
\begin{align*}
&-\frac{\psi}{2L} u^2_i(t_i^k)\tau^3+ \big (0.5  \Delta u_{i,j}(t_i^k) - \frac{3\psi}{2L}u^2_i(t_i^k)+
\\&-\frac{3\psi}{2L} v_{i}(t_i^k) u_i(t_i^k) \big ) \tau^2
+\Big (\Delta u_{i,j}(t_i^k)-\frac{3\psi}{L}v_i(t_i^k)u_i(t_i^k)
\end{align*}
\begin{align*}
&+  ( \Delta v_{i,j}(t_i^k)-\frac{\psi}{L} v^2_i(t_i^k)-\frac{\psi}{L} u_i(t_i^k)x_{i}(t_i^k)) \Big)\tau\\
 &+\mathcal{C}_{i,4}(t_i^k,u_i(t_i^k))=0,
\end{align*}
where $\Delta v_{i,j}(t_i^k)=v_{j}(t_{j}^k)-v_i(t_i^k)$ and $\Delta u_{i,j}(t_i^k)=u_{i}(t_{j}^\textmd{last})-u_i(t_i^k)$.  The least positive root   is denoted  as $\tau_{i,4}$ and we define $t^k_{i,4}=t_i^k+\tau_{i,4}$. Moreover, due to the added term in \eqref{cbf_modified_4}, it follows that $\tau_{i,4}>T_d$.  The case of having all roots negative corresponds to the scenario that the constraint \eqref{eq_15} will not be violated, hence $t^k_{i,4}=\infty$.

\subsection{Self-Triggered Scheme}

First, it should be noted that in the previous section, the time instants $t^k_{i,j}$, $j=1,\dots,4$  are obtained based on the safety constraints \eqref{h1} and \eqref{h2}, as well as the vehicle state limitations \eqref{h3}. However, this can compromise the optimal performance of CAVs passing the CZ. This is because it may happen that the CAV acceleration is forced to remain constant for a long period due to the fact that no safety constraints or vehicle state limit violation occurs, whereas, as per  \cite{xiao2021bridging}, the optimal acceleration trajectory of the CAV in fact changes linearly. Therefore, in order to avoid this issue and reinforce the optimal acceleration trajectory, one can impose a maximum allowable inter-event time, denoted as $T_{\max}$. To accomplish this, we can define
\begin{align}
    t^k_{i,\min}=\min\{t^k_{i,1},t^k_{i,2},t^k_{i,3},t^k_{i,4},t_i^k+T_{\max} \}.
\end{align}

The next update time instant for CAV $i$, i.e. $t_i^{k+1}=t_i^\textmd{next}$ should now be calculated. Towards this goal, if $t^k_{i,\min} \leq \min(t_{i_p}^\textmd{next},t_{j}^\textmd{next})$
which corresponds to the case that the next update time instant of CAV $i$ should occur before the next update of the preceding vehicle $i_p$ or the CAV $j$. Then, we can set $t_i^{k+1}=t_i^\textmd{next}=t_{i,\min}$.

The only remaining case to be considered is when $t^k_{i,\min}> \min(t_{i_p}^\textmd{next},t_{j}^\textmd{next})$, which corresponds to either CAVs $i_p$  or $j$ updating its control input sooner than CAV $i$, hence CAV $i$ does not have access to their updated control input. Consequently, checking the constraints \eqref{eq_14} and \eqref{eq_15}  is not anymore valid. In this case, $t_i^\textmd{next}= \min(t_{i_p}^\textmd{next},t_{j}^\textmd{next})+T_d$ which implies that the $i$-th CAV next update time will be immediately  after the update time of  CAVs $i_p$  or $j$ with minimum time-interval $T_d$. 

To summarize, we have
\begin{align}\label{t_next}
    t_i^{\textrm{next}}=\left \{ \begin{array}{ll}
        t_{i,\min}, & t_{i,\min}\leq \min(t_{i_p}^\textmd{next},t_{j}^\textmd{next}),\\
        \min(t_{i_p}^\textmd{next},t_{j}^\textmd{next})+T_d, & \textrm{otherwise},
    \end{array} \right.
\end{align} 
Finally, in order to have $(t_i^k~\textrm{mod}~T_d)=0$, we set $t_i^\textmd{next}=\lfloor \frac{t_i^\textmd{next}}{T_d} \rfloor \times T_d$. It should be noted that the case of  $t_i^{\textrm{next}}=t_j^{\textrm{next}}$ or $t_i^{\textrm{next}}=t_{i_p}^{\textrm{next}}$  corresponds to having identical next update times for CAV $i$ and CAV $j$ or CAV $i_p$ so that they need to solve their QPs at the same time instant. However, in order for CAV $i$ to solve its QP at the time instant $t_i^{k+1}=t_i^\textmd{next}$, it requires the updated control input of CAV $j$ or CAV $i_p$, i.e. $u_j(t_i^{k+1})$ or $u_{i_p}(t_i^{k+1})$; this is practically not possible. In order to remedy this issue, whenever  $t_i^{\textrm{next}}=t_j^{\textrm{next}}$ or $t_i^{\textrm{next}}=t_{i_p}^{\textrm{next}}$, CAV $i$ solves its QP at $t_i^{k+1}$ by  substituting  $u_M$ instead of $u_j(t_i^{k+1})$ and  $u_{i_p}(t_i^{k+1})$ in \eqref{cbf_modified_3} and \eqref{cbf_modified_4}. This indeed corresponds to considering the worst case in $\nu_{i,3}(T_d,t)$ and $\nu_{i,4}(T_d,t)$. Moreover, as calculating the next update time $t_i^{k+2}$ also depends on $u_j(t_i^{k+1})$ and  $u_{i_p}(t_i^{k+1})$, CAV $i$ acts similar to the time-triggered case with assigned  $t_i^{k+2}=t_i^{k+1}+T_d$. Then, at the next time instant $t_i^{k+2}$, CAV $i$ can obtain the updated control inputs of CAV $j$ and CAV $i_p$ from the coordinator and follows the proposed self-triggered scheme.

\section{Simulation Results} \label{Sec_sim}
The simulation is conducted using MATLAB for the merging problem shown in Fig. \ref{Fig:Merge_Problem} where CAVs arrive according to Poisson arrival processes with an arriving rate. The initial speed $v_{i}(t_{i}^{0})$ is also randomly generated with a uniform distribution over $[15 \textnormal{m/s}, 20\textnormal{m/s}]$ at the origins $O$ and $O^{\prime}$, respectively.  The parameters in the QP problem \eqref{QP} are $L=400 \textrm{m}$, $\psi=1.8\textmd{s}$, $l=0$, $g=9.81 \textmd{m}/\textmd{s}^2$, $v_{\max}=30 \textmd{m}/\textmd{s}$, $v_{\min}= 0 \textmd{m}/\textmd{s}$, $c_3=10$, $c_d=0.6$, $c_a=0.5$, and $\rho=1$.

In addition to a simple surrogate $L_2$-norm ($u^2$) model, the following energy consumption model is used for comparison and performance analysis between different approaches  \cite{kamal2012model} :
$
f_v(t)=f_{\textrm{cruise}}(t)+f_{\textrm{accel}}(t), 
f_{\textrm{cruise}}(t)= \omega_0+\omega_1v_i(t)+\omega_2v^2_i(t)+\omega_3v^3_i(t), 
f_{\textrm{accel}}(t)=(r_0+r_1v_i(t)+r_2v^2_i(t))u_i(t)
$
where we used typical values for parameters $\omega_1,\omega_2,\omega_3,r_0,r_1$, and $r_2$ as reported in \cite{kamal2012model}.

The time-triggered scheme with $T=0.05 \textmd{s}$ and the proposed self-triggered scheme with $T_d=0.05\textmd{s}$ and different $T_{\max}$ are implemented in MATLAB where \textsc{quadprog} is used for solving QP \eqref{QP2}.  Table \ref{Table I} shows the summary of the results for 28 different scenarios corresponding to 3 different approaches under the same traffic flow and initial velocity profile. The performance of the proposed self-triggered scheme under different values of $T_{\max}\in\{0.5,1,1.5,2\}$ shows that the number of communication between the CAVs and coordinator is significantly reduced (more than 78\%). Moreover, as observed for the $T_{\max}=0.5$, one can still achieve comparable average fuel consumption and    $L_2$-norm ($u^2$) performance while the number of communications is reduced to $20.46\%, 19.5\%, 20.4\%$, and $21.8\%$ for different values of $\alpha\in\{0.1,0.25,0.4,0.5\}$. In addition, as expected, as $T_{\max}$ is increased, the number of events is further reduced by the price of having larger $L_2$-norm ($u^2$) as the CAVs accelerations remain constant for a longer interval and had sharper changes during the update.

Furthermore, in order to further investigate the effect of modified CBFs \eqref{cbf_modified_1}, \eqref{cbf_modified_2}, \eqref{cbf_modified_3}, and \eqref{cbf_modified_4} on different performance measures, the time-triggered scheme using modified CBFs is simulated for different values of $\alpha$. As it is shown in Table \ref{Table I}, the performance measures are slightly modified in comparison with the time-triggered scheme using the original CBFs \eqref{eq_12}-\eqref{eq_15} and hence the proposed modification does not add any conservativeness to the coordination of CAVs. Finally, as shown in Table \ref{Table I}, the self-triggered scheme with $T_{\max}$ has better performance in terms of reducing the number of communication while having similar average travel time, average fuel consumption and    $L_2$-norm ($u^2$) performance.

\begin{table*}\scriptsize
        \centering
        \begin{tabular}{|c|c|c|c|c|c|c|c|c|}
            \cline{1-9}
             &Item & \multicolumn{4}{|c|}{Self-Triggered} & Event-Triggered  &Time-Triggered & Time-Triggered\\
             & & \multicolumn{4}{|c|}{} & \cite{event_triggered} & Modified CBF & \\
            \cline{2-9}
            & $T_{\max}$ & $0.5$ & $1$ & $1.5$ & $2$ & $s_v=0.5, s_x=1.5$ & $T_s=T_d = 0.05$ & $T_s = 0.05$\\
        \hline  
        \multirow{5}{*}{{$\alpha=0.1$ }} & Ave. Travel time & 19.5 & 19.48 & 19.48& 19.49& 19.61 & 19.5 &19.42\\
        \cline{2-9}
        & Ave. $\frac{1}{2} u^2$ & 4.27 & 5.00 & 5.93 & 7.2 & 4.45 & 3.37 & 3.18\\
        \cline{2-9}
        & Ave. Fuel consumption & 31.86 & 32.21 & 32.64 & 33.23 & 31.77 & 31.32 &31.61\\
        \cline{2-9}
        &Total Number of & 20.46\% &  11.9\%  & 10.87\% & 10.32\% & 50\% & 100.5\% &100\% \\
                & Communication & (7252)&  (4218) & (3854) &(3658)& (17853) & (35636) &(35443) \\
        \cline{2-9}
         \hline
         \hline
        \multirow{5}{*}{{$\alpha=0.25$ }} & Ave. Travel time & 15.57 & 15.56 & 15.57 &15.62& 15.82 &15.58&15.44\\
        \cline{2-9}
        & Ave. $\frac{1}{2} u^2$& 14.33 & 15.10 & 15.68 & 16.68 & 13.93 &13.38 &13.34\\
        \cline{2-9}
        & Ave. Fuel consumption & 54.45 & 53.51  & 52.57 & 52.94 & 52.12 & 54.17 &55.81 \\
        \cline{2-9}
        &Total Number of &19.5\%  & 13.68\%  & 12.34\%  & 12.72\%  & 51\% &100.9\% &100\% \\
                &Communication & (5495) & (3857) & (3479) & (3588) & (14465)&(28461) & (28200)\\
        \hline
        \hline
                \multirow{5}{*}{{$\alpha=0.4$ }} & Ave. Travel time & 15.15 &  15.15 & 15.18  &15.2& 15.4 & 15.16 &15.01 \\
        \cline{2-9}
        & Ave. $\frac{1}{2} u^2$& 18.5 & 19.32 & 19.73 & 20.36 & 18.04 & 17.64 & 17.67\\
        \cline{2-9}
        & Ave. Fuel consumption & 55.23 & 53.35 & 52.67 & 52.95& 53.15 &54.93 &56.5\\
        \cline{2-9}
        &Total Number of   & 20.4\%  & 14.85\%  & 13.69\%  & 13.60\%  & 54\%  & 101.0 \% & 100\% \\
                & Communication    &  (5591) & (4071) & (3754) & (3727) & (14089) & (27695)  & (27412)\\
        \hline   
        \hline
                \multirow{5}{*}{{$\alpha=0.5$}} & Ave. Travel time & 14.79 & 14.79 & 14.82 & 14.89& 15.09 & 14.8 &14.63 \\
        \cline{2-9}
        & Ave. $\frac{1}{2} u^2$ &25.5 & 25.84  & 26.43 & 27.5 & 24.94 &24.86 &25.08\\
        \cline{2-9}
        & Ave. Fuel consumption & 55.5 & 53.15 & 52.9 & 53.45 	& 53.65 & 55.5 &56.93 \\
        \cline{2-9}
        &Total Number of  & 21.8\%    & 16.7\%  & 15.09\%  &  15.17\%  & 51\%  & 101.1\% & 100\%  \\
                &Communication & (5841) &  (4322) &  (4034) &  (4054) &  (13764) & (27033) & (26726) \\
        \hline        
        \end{tabular}
        \caption{CAV metrics under self-triggered and time-driven control.  }
        \label{Table I}
\end{table*}

\section{Conclusion}\label{Sec_Con}
A self-triggered safe coordination control of CAVs in conflict areas of a traffic network is developed with the main aim to reduce the number of communication between CAVs and the coordinator. Toward this, modified CBF constraints are obtained to assure the minimum inter-event interval and the event time instants are obtained by checking when the CBF conditions are violated. Finally, the maximum allowable inter-event time is defined to enforce the CAVs follow the optimal solution as much   as possible. As the future work, the robustness of the proposed self-triggered scheme with respect to CAVs dynamic uncertainty as well as environmental disturbances will be investigated. Moreover, a full comparison between the event-triggered and self-triggered schemes is a topic on ongoing research.
\bibliographystyle{IEEEtran}
\bibliography{Report}

\end{document}